\PassOptionsToPackage{prologue,table}{xcolor}
\documentclass[manuscript,screen,nonacm]{acmart}  %
\acmSubmissionID{XXXX}

\PassOptionsToPackage{prologue,table,xcdraw}{xcolor}
\usepackage{xargs}
\usepackage{booktabs} %
\usepackage{tabularx}
\usepackage{caption}
\usepackage{nicematrix}
\usepackage{float}
\usepackage{mathtools}
\usepackage{amsmath,amsfonts}
\usepackage{bm}
\usepackage{microtype}
\usepackage{units}
\usepackage{subcaption}
\usepackage{overpic}
\usepackage{cancel}
\usepackage{wrapfig}
\usepackage{placeins}
\usepackage{empheq}
\usepackage[normalem]{ulem}
\usepackage[export]{adjustbox}
\usepackage[vlined,boxed,ruled]{algorithm2e} %
\usepackage{tikz}
\usepackage{xstring}
\usepackage{xr}
\usepackage{multirow}
\usepackage[capitalize,noabbrev]{cleveref} %
\usepackage{url} %
\usepackage[utf8]{inputenc}
\usepackage{wrapfig}
\usepackage{listings}
\usepackage{epigraph}
\usepackage{enumitem}
\usepackage{csquotes}
\usepackage[most]{tcolorbox}
\usepackage{etoolbox}

\patchcmd{\quote}{\rightmargin}{\leftmargin 3em \rightmargin}{}{}

\setlist[itemize]{noitemsep, nolistsep, leftmargin=*}
\setlist[enumerate]{noitemsep, nolistsep, leftmargin=*}

\SetKw{Or}{or}
\SetKw{And}{and}
\SetStartEndCondition{ }{}{}%
\SetKwProg{Fn}{def}{\string:}{}
\SetKwFunction{Range}{range}%
\SetKw{KwTo}{in}
\SetKwFor{For}{for}{\string:}{}%
\SetKwFor{While}{while}{:}{fintq}%
\SetKwIF{If}{ElseIf}{Else}{if}{:}{elif}{else:}{}%
\SetKwComment{Comment}{\# }{}
\SetKw{Continue}{continue}
\AlgoDontDisplayBlockMarkers
\SetAlgoNoEnd
\SetAlgoNoLine
\DontPrintSemicolon

\SetAlFnt{\small}
\SetAlCapFnt{\small}
\SetAlCapNameFnt{\small}
\SetAlCapHSkip{0pt}
\IncMargin{-\parindent}

\definecolor{maroon}{cmyk}{0, 0.87, 0.68, 0.32}
\definecolor{halfgray}{gray}{0.55}
\definecolor{ipython_frame}{RGB}{207, 207, 207}
\definecolor{ipython_bg}{RGB}{247, 247, 247}
\definecolor{ipython_red}{RGB}{186, 33, 33}
\definecolor{ipython_green}{RGB}{0, 128, 0}
\definecolor{ipython_cyan}{RGB}{64, 128, 128}
\definecolor{ipython_purple}{RGB}{170, 34, 255}

\lstdefinelanguage{iPython}{
    morekeywords={access,and,break,class,continue,def,del,elif,else,except,exec,finally,for,from,global,if,import,in,is,lambda,not,or,pass,print,raise,return,try,while},%
    morekeywords=[2]{abs,all,any,basestring,bin,bool,bytearray,callable,chr,classmethod,cmp,compile,complex,delattr,dict,dir,divmod,enumerate,eval,execfile,file,filter,float,format,frozenset,getattr,globals,hasattr,hash,help,hex,id,input,int,isinstance,issubclass,iter,len,list,locals,long,map,max,memoryview,min,next,object,oct,open,ord,pow,property,range,raw_input,reduce,reload,repr,reversed,round,set,setattr,slice,sorted,staticmethod,str,sum,super,tuple,type,unichr,unicode,vars,xrange,zip,apply,buffer,coerce,intern},%
    sensitive=true,%
    morecomment=[l]\#,%
    morestring=[b]',%
    morestring=[b]",%
    morestring=[s]{'''}{'''},%
    morestring=[s]{"""}{"""},%
    morestring=[s]{r'}{'},%
    morestring=[s]{r"}{"},%
    morestring=[s]{r'''}{'''},%
    morestring=[s]{r"""}{"""},%
    morestring=[s]{u'}{'},%
    morestring=[s]{u"}{"},%
    morestring=[s]{u'''}{'''},%
    morestring=[s]{u"""}{"""},%
        literate=
    {^}{{{\color{ipython_purple}\^{}}}}1
    {=}{{{\color{ipython_purple}=}}}1
    {+}{{{\color{ipython_purple}+}}}1
    {*}{{{\color{ipython_purple}$^\ast$}}}1
    {/}{{{\color{ipython_purple}/}}}1
    {+=}{{{+=}}}1
    {-=}{{{-=}}}1
    {->}{{{->}}}1
    {*=}{{{$^\ast$=}}}1
    {/=}{{{/=}}}1,
    literate=
    *{-}{{{\color{ipython_purple}-}}}1
     {?}{{{\color{ipython_purple}?}}}1,
    identifierstyle=\color{black}\ttfamily,
    commentstyle=\color{ipython_cyan}\ttfamily,
    stringstyle=\color{ipython_red}\ttfamily,
    keepspaces=true,
    showspaces=false,
    showstringspaces=false,
    rulecolor=\color{ipython_frame},
    frame=single,
    frameround={t}{t}{t}{t},
    numbers=left,
    numberstyle=\tiny\color{halfgray},
    backgroundcolor=\color{ipython_bg},
    basicstyle=\scriptsize,
    keywordstyle=\color{ipython_green}\ttfamily,
    numbersep=8pt,
    captionpos=b,
    breaklines=true,
    breakatwhitespace=true,
    showstringspaces=false,
    showtabs=false,
    tabsize=1,
}

\definecolor{ListBGColor}{rgb}{0.95,0.95,0.95}
\definecolor{ListCommentColor}{rgb}{0.3,0.5,0.3}
\definecolor{ListKeywordColor}{rgb}{0.2,0.2,0.7}
\definecolor{ListStringColor}{rgb}{0.6,0.1,0.1}
\definecolor{ListNumberColor}{rgb}{0.4,0.4,0.4}
\definecolor{ListIdentifierColor}{rgb}{0.1,0.1,0.1}
\def\commentType{0}

\ifnum\commentType=0
    \newcommandx{\customComment}[3]{}
    \newcommandx{\customTODO}[3]{}
\fi
\ifnum\commentType=1
    \usepackage[
    type=upperleft,
    unit=in
    ]{fgruler}
    \newcommandx{\customComment}[3]{\textbf{\scriptsize \color{#2} #1: #3}}
    \newcommandx{\customTODO}[3]{\textbf{\scriptsize \color{#2} #1: #3}}
\fi
\ifnum\commentType=2
    \usepackage[
    type=upperleft,
    unit=in
    ]{fgruler}
    \usepackage{pdfcomment}
    \newcommandx{\customComment}[3]{\pdfcomment[icon=Comment,opacity=0.5,color=#2,author=#1]{#3}}
    \newcommandx{\customTODO}[3]{\pdfcomment[icon=Note,opacity=0.5,color=#2,author=#1]{#3}}
\fi
\ifnum\commentType=3
    \usepackage[
    type=upperleft,
    unit=in
    ]{fgruler}
    \usepackage{pdfcomment} %
    \usepackage{todonotes} %
    \usepackage{silence}
    \newcommandx{\customComment}[3]{\todo[color=#2!40,size=\small]{\textbf{#1:} #3}}
    \newcommandx{\customTODO}[3]{\todo[color=#2!40,size=\small]{\textbf{#1:} #3}}
\fi

\let\originalleft\left %
\let\originalright\right %
\renewcommand{\left}{\mathopen{}\mathclose\bgroup\originalleft} %
\renewcommand{\right}{\aftergroup\egroup\originalright} %

\definecolor{amber}{rgb}{1.0, 0.49, 0.0}
\definecolor{darkgreen}{rgb}{0.0, 0.5, 0.0}
\definecolor{darkblue}{rgb}{0.0, 0.0, 0.5}
\definecolor{darkred}{rgb}{0.5, 0.0, 0.0}
\definecolor{bloodred}{rgb}{0.616, 0.133, 0.208}
\definecolor{forestgreen}{rgb}{0.133, 0.616, 0.541}
\definecolor{babyblue}{RGB}{91, 206, 250}
\definecolor{babypink}{RGB}{245, 169, 184}

\newcommandx{\All}[1]{\customComment{All}{red}{#1}}
\newcommandx{\ana}[1]{\customComment{AD}{babyblue}{#1}}
\newcommandx{\moira}[1]{\customComment{MW}{forestgreen}{#1}}

\newcommandx{\thesis}[1]{\customComment{\textsc{Thesis}}{bloodred}{#1}}
\newcommandx{\TODO}[1]{\customTODO{TODO}{red}{#1}}
\newcommandx{\todo}[1]{\customTODO{TODO}{red}{#1}}

\newcommand{\REMOVE}[1]{} %
\newcommand{\REPLACE}[2]{#2} %

\def\equationautorefname~#1\null{%
  Equation~(#1)\null%
}

\newcommand{\socwork}[1]{social doing}
\newcommand{\SocWork}[1]{Social Doing}

\let\oldcdot\cdot
\renewcommand{\cdot}[0]{{\;\oldcdot\;}}

\DeclareMathAlphabet{\mathmybb}{U}{bbold}{m}{n}

\definecolor{cartoPrismTeal}{rgb}{0.21960784 0.65098039 0.64705882}
\definecolor{cartoPrismOrange}{rgb}{0.88235294 0.48627451 0.01960784}
\definecolor{cartoPrismGreen}{rgb}{0.45098039 0.68627451 0.28235294}
\definecolor{cartoPrismRed}{rgb}{0.8 0.31372549 0.24313725}
\definecolor{cartoPrismPurple}{rgb}{0.58039216 0.20392157 0.43137255}

\definecolor{mathematicaBlue}{rgb}{0.38, 0.51, 0.71}
\definecolor{mathematicaOrange}{rgb}{0.88, 0.61, 0.14}
\definecolor{mathematicaGreen}{rgb}{0.56, 0.69, 0.19}
\definecolor{mathematicaRed}{rgb}{0.92,0.39, 0.21}
\definecolor{mathematicaPurple}{rgb}{0.53, 0.47, 0.7}

\newtcolorbox[auto counter]{opportunitybox}{ %
  enhanced,
  colback=cartoPrismTeal!15,
  colframe=gray,
  boxrule=0pt,
  arc=6pt,
  left=6pt,
  right=6pt,
  top=4pt,
  bottom=4pt,
  fuzzy shadow={0pt}{0pt}{-1pt}{0.5pt}{gray},
  beforeafter skip balanced={0.5em},
  after skip={0pt},
  before upper=\textbf{Opportunity~\thetcbcounter: }\ignorespaces%
}
\usepackage{wrapfig}

\begin{document}

\title[Synthetic Sociality: How Generative Models Privatize the Social Fabric]{Synthetic Sociality: How Generative Models Privatize the Social Fabric} %

\author{Ana Dodik}
\email{anadodik@mit.edu}
\orcid{0000-0003-4391-8877}
\affiliation{%
  \institution{MIT CSAIL}
  \country{USA}
  \city{Cambridge, MA}
}

\author{Moira Weigel}
\email{weigel@fas.harvard.edu}
\orcid{0009-0006-5795-0754}
\affiliation{%
  \institution{Harvard University}
  \country{USA}
  \city{Cambridge, MA}
}

\begin{abstract}
        We put forth a critical theoretical framework for analyzing generative models both descriptively and normatively.
        Our thesis is that generative models automate the production not only of intellectual labor or intelligence, but of a broader set of human social capacities we name ``\socwork{}.'' 
        We do this by historicizing the commodification of sociality in the digital economy, leading to the availability of social data as the precondition for generative models.
        We elaborate our definition of ``\socwork{}'' by drawing a distinction between ``use'' and ``exchange'' sociality and further differentiate between the ways that generative models either substitute for or mediate existing social relations and processes.
        We then turn to existing empirical research on how people use generative model-based products and the effects that their use has upon them.
        In this, we introduce the concept of Synthetic Sociality, a social reality in part fabricated by Silicon Valley's privately owned and undemocratically governed generative models.
        Lastly, we offer a normative analysis based on our findings and framework, and discuss future design opportunities.
\end{abstract}



\maketitle

\setlength\epigraphwidth{.75\linewidth}
\epigraph{\emph{Again, by [artificial general intelligence], we mean a highly autonomous system that outperforms humans at most economically valuable work.} --- \citet{openai2025structure}}{}

\section{Introduction}\label{s:intro}

Prompted by their widespread deployment and adoption of generative statistical models, critical theory, and its emerging subfield of so-called ``critical artificial intelligence studies,'' is attempting to reckon with their place in society.
Thus far, most theories in the materialist tradition treat generative models as forms of deskilling and automation of mental labor,  intelligence, or language~\cite{pasquinelli2023eye, tenen2024literary, biondi2023specter, weatherby2025language}.
Through the historical lens of labor automation, they conclude, generative models are not special or salient in history. %
Yet, certain use-cases and impacts are unprecedented in kind or magnitude.
Prominent ones include generating messages to loved ones in our stead, proliferating synthetic media on social networks at industrial scale, and having us perceive a romantic relationship with a virtual avatar.
An emergent through line is that, in exchange for a monthly subscription fee, generative models promise to displace social roles hitherto assumed only by other social beings.
Therefore, to explain their impacts, critical theory must historicize why generative models, despite being ontologically excluded from the category, are successfully deployed in lieu of social beings.

We build on existing scholarship, departing in two ways.
First, we focus on generative statistical models and their deployments, not all of artificial intelligence~\cite{pasquinelli2023eye} or only language modeling~\cite{tenen2024literary, weatherby2025language}.
We include language, image, video, and audio generative models and exclude things like robot planning, image classification, and protein folding prediction.
Second, and more importantly, we assert that these models automate \emph{not only} mental labor, ``social intelligence'', or ``general intellect''~\cite{babbage1832economy, pasquinelli2023eye}, intelligence immanent to language and its recombinability~\cite{tenen2024literary,weatherby2025language}, or artistic production~\cite{epstein2023art}.
Rather, we provide a materialist theory of generative models as automating the production of our social fabric.

Specifically, generative models automate a broader set of capacities that we call \emph{\socwork{}}, defined as what we exert to build a social connection with another.
We elaborate our definition of \socwork{} by drawing a distinction between \emph{use \socwork{}} and its commodity form of \emph{exchange \socwork{}}.
Building on existing critical scholarship on the sociality's relationship to the political economy~\cite{fraser2016crisis}, we historicize its commodification, focusing on its most recent evolution under the digital economic mode.
This tireless conquest of the digital commodity form over our social fabric birthed the preconditions for generative models to come and automate its production (\S\ref{s:sociality}): generative models can only automate \socwork{} because of data with social characteristics shared on digital platforms.
Thus, our theory views generative models as automating \socwork{} via the accumulation of dead labor and of \emph{dead \socwork{}}, and their deployments as expressions of the fungibility and exchange character of digital sociality.

We differentiate between two kinds of automation: substitutive and mediative.
Substitutive automation (\S\ref{ss:direct}), e.g., by a companion chatbot, means completely replacing the \socwork{} of a human.
Mediative automation (\S\ref{ss:indirect}), e.g., image generators, only partially replaces or supplements the \socwork{} of the model's user---in addition to replacing the \socwork{} they could have themselves conceivably put into a social tie, the model also gives them capacities of the \emph{dead \socwork{}} contained inside it.
We argue our theory through an interpretative textual analysis of primary sources, including empirical research, media reporting, and corporate communications.
In it, we focus on people's use of generative model-based products, like chatbots or image generators, and the effects this use has on them (\S\ref{ss:direct}, \S\ref{ss:indirect}).

Finally, we argue that many ethical considerations around generative models lay downstream from the automation and privatization of \socwork{}.
In Section~\ref{s:analysis}, we draw conclusions from our theoretical foundation and offer design opportunities for future generative models.
First, we argue that the commodity fetishism inherent to automation severs the connection to the living \socwork{} in the training data and obfuscates questions of economic attribution and cultural bonds and historicity (\S\ref{ss:fetish}).
Next, we discuss the dialectic nature of generative \socwork{} and ask whether there are scenarios where automation of \socwork{} might be beneficial (\S\ref{ss:regime}), before briefly discussing how the cultural impacts of automation of \socwork{} (\S\ref{ss:regime}).
Finally, we explore how at-will access to social capacities without the need to employ humans allows the owners and users of models to alter our perceptions of self, our relationships, and society at large, and how private ownership ensues a lack of transparency and democratic control over the automated manufacture of our social fabric (\S\ref{ss:fever}).
Silicon Valley's privately owned and governed models alter our social fabric into a partially generated hollow imitation---a \emph{simulacrum}---which we dub Synthetic Sociality.(\S\ref{ss:generativesociality}).

In summary, our contributions are:
\begin{itemize}
    \item A materialist theory of how generative models automate and privatize the manufacture of our social fabric.
    \item The concept of \emph{\socwork{}} as a theoretical foundation and a tracing of its historical evolution.
    \item A textual analysis of primary sources for generative models' involvement and impact on social roles.
    \item A normative analysis of the automation and privatization of \socwork{} and design opportunities for future models.
\end{itemize}

\section{Is Intelligence All There Is?}\label{s:intellect}

No discussion of generative models, language, image, or otherwise, can escape veering into a discussion on intelligence.
Generative models do partially fall under the vast academic umbrella of artificial intelligence and they do continue the long history of automation of intellectual tasks such as mathematical derivations, code generation, and information retrieval.
As \citet{tenen2024literary} points out, language models fit squarely into the tradition of tools meant to automate away parts of the writing process, and one could likely make similar arguments for visual art, tracing the histories of photography and computer graphics technologies.
Similarly, \citet{weatherby2025language} explores how automation of linguistic and intellectual capacities (e.g., making metaphors, doing math), shapes what we consider human.
Zooming out, \citet{pasquinelli2023eye} and \citet{biondi2023specter} frame artificial intelligence (AI) as \emph{alienation} and \emph{reification} of \emph{general intellect}, and/or \emph{social knowledge} from workers into machinery, tracing it back to the industrial revolution.
Therefore, they conclude, generative models are not salient in history because the automation of intelligence is not new---they present a difference of quantity, not quality.

However, generative models are at the center of a societal upheaval, be it through stupendous capital investments, the ensuing ecological impact, or the loud and decisive backlash from parts of the population.
Claiming, as prior work does, that generative models are par for the course---just more of the same---stops us short of asking the interesting question: what makes them different?

To understand the hype and the pushback alike, we must focus on what current models excel at.
Language model generations are used for humans to transmit information without needing to talk to or interact with each other: instead of finding words within ourselves, we exchange them for a machine's pastiche.
At home, chatbot companions offer an emulated feeling of companionship without a human offering it.
Culturally, generated artifacts stripped bare of the individual and cultural historical context required to make art are nonetheless paraded as art: beautiful visuals with nothing to say.
As noted by \citet{kreminski2024dearth}, their users input little information compared the richness of their outputs.
This certainly implies automation, but here we must pause and ask. 
Is intelligence really all there is to self-expression, companionship, communication, culture and art?
All indeed require intelligence, but the key feature in all is that they are expressions of our sociality.
They are important not because of their utilitarian purpose as constructs of intellect, but because they are relational to another.
They are important because they exist as a reflection of ourselves and others and the bonds in between.
Thus, a critical theory of generative models must explicitly account for automating sociality.

The cultural and societal aspect of generative models has been noted by \citet{farrell2025culturalai}, who compare them to societal institutions tasked with exchanging information, such as markets, bureaucracies or democracies.
Ultimately though, this work still focuses on information exchange, and, moreover, takes an uncritical approach.
It misses the cruicial point that societal institutions are living organisms, reliant on living sociality and evolving culture, whereas generative models reify ``dead'' sociality into a privately owned technology.
In contrast, our analysis views generative models critically as machinery that fabricates sociality, as a manufacture simulacra of that which we use to connect.

\section{Background}\label{s:background}

The history of computation is, in many ways, the history of the division of labor.
Extending Adam Smith to the realm of mental labor, Charles Babbage imagined the first computer~\cite{smith1947wealth, babbage1832economy}.
He was inspired by De~Prony, who divided the mental labor for computing logarithms into atomized steps to be performed by deskilled factory workers~\cite{deprony1824notice}; hearing of this, Babbage suggested that the deskilled labor could be replaced by machines, ones which we now call computers~\cite{babbage1832economy, grattanguinness2003deprony,pasquinelli2023eye, whittaker2023origin}.
Computers thus became as machinery---fixed capital to be privately owned, to self-valorize~\cite{smith1947wealth, babbage1832economy}.
Following this line of reasoning, prior critical theory has identified ``artificial intelligence'' as yet another form of automation and deskilling of intellectual labor \cite{pasquinelli2023eye, tenen2024literary}.

Machinery is a fundamental constitutive element of the capitalist mode of production.
Fiduciary duty to the profit motive often requires lowering costs, but labor comes with a lower bound on its price:~the lowest wage compatible with the so-called \emph{common humanity}, one necessary for the worker's survival~\cite{smith1947wealth}.
At some point, further decreasing production costs necessitates automation~\cite{mackenzie1984marx}.
Once machinery exists, new workers become but its components, and we forget of the labor it imitates, the labor that designed and constructed it.
These workers do not see the profits of the machine; the wealth is privately owned.
We refer to this forgotten labor as \emph{dead labor}, and to the idea of a machine as a magical object, a force of nature whose capacities belong to it instead the workers who made it, as \emph{fetishism}.

During the era of \emph{multinational capitalism}, labor in the Global North became immaterial and thus more reliant on sociability, commodification expanded, and the cultural and economic spheres began collapsing into one another~\cite{lazzarato1996immaterial, mandel1975late, jameson1991postmodernism}.
Recently, scholars have theorized the emergence of a new, digital era of the political economy, characterized by quantitative easing and the emergence of data as a commodity~\cite{srnicek2016platform}.
Interpretations of the datafied economy vary, some starting from a labor perspective~\cite{terranova2000free, jarrett2014womens, ptak2014wages}, others from capitalism's relationship to colonialism~\cite{morozov_critique_2022, mejias2024data}, and others yet feudalism~\cite{varoufakis2023technofeudalism}.
Our analysis should be compatible with any of these perspectives; building upon \citet{lukacs1923reification}, what matters for us is that data is a commodity produced by people, and sold and traded in the market economy.

The fundamental part of any commodity, data included, is the divorcing of its use value (how valuable people find a good or service) from its exchange value (how much the free market is willing to pay for it)~\cite{smith1947wealth}.
When society uses commodity exchange as its exclusive value signifying mechanism, we say that the use value of a good or service is subsumed by its exchange value as a commodity form.
Consequentially, any two of the same commodity form become interchangeable, i.e., fungible, on the free market: the price of a family heirloom is not increased by its emotional value and water is worth less than diamonds~\cite{smith1947wealth}.
In the datafied economy, while an online purchase gives convenient access to items, the data about the purchase carries added exchange value for the platform~\cite{srnicek2016platform, mejias2024data,varoufakis2023technofeudalism,ptak2014wages}.

As we will see, datafication has had a profound effect on human sociality, one that makes generative models a logical conclusion of its evolution.
In Section~\ref{s:sociality}, we will explore the subsumption of sociality into commodity relations.

\section{Automated Privatized Sociality}\label{s:sociality}

To think the role of generative models in society, one must look at the underlying economic base that engenders them.
We posit that the automation of human social relations via generative models parallels the automation of mental labor: once commodification of social ties is all-permeating and the fungibility of sociality is instituted, our social ties are ready for automation.
We begin by introducing theoretical tools for discussing the commodification of social ties (\S\ref{ss:socwork}) and apply it to a historical account (\S\ref{ss:fungsoc}), before returning to generative models (\S\ref{ss:generativesociality}) as supported via our literature survey (\S\ref{ss:direct}, \S\ref{ss:indirect}).
Lastly, we offer a normative analysis via our theory and discuss future design opportunities (\S\ref{s:analysis}).

\subsection{\SocWork{} and Content}\label{ss:socwork}

We now introduce a theoretical framework and establish a common language to explain this historical process.
We use \emph{social relation} in the sociological sense as the concept referring to any connection between people---we use \emph{social tie} as a synonym. Social ties weave together into the \emph{social fabric}, the grand sum of all social relations.
The production of social ties consumes and transforms a resource we dub \emph{\socwork{}}.
It is what makes social ties, what animates the loom that weaves our social fabric.
It is \emph{not} a quantity to be exchanged between social beings; rather social beings can choose to use it to create the social ties that make up their life.
Some scholars would assert all social doing is labor power~\cite{lazzarato1996immaterial,terranova2000free,federici1975wages,fortunati1995arcana}.
We intentionally steer clear from this form of economic reductionism, viewing \socwork{} as a broader category. %
Instead, following \citet{lukacs1923reification} we find commodification a necessary and sufficient condition for discussing social relations.
Therefore, \socwork{} is not \emph{necessarily} labor power, rather a resource which sometimes becomes a commodity.
We discuss this choice in the context of its alternatives~\cite{fraser2016crisis,federici1975wages, fortunati1995arcana, jarrett2014womens}, below.

Social doing is inseparable from the economy: not only is labor power a form of \socwork{}, but \citet{fraser2016crisis} teaches us capital transforms all relations, regardless of proximity to capital.
A key influence is the commodification of \socwork{}.
Thus, the use-exchange dialectic is particularly pertinent to \socwork{}, which itself has bifurcated into \emph{use \socwork{}} and \emph{exchange \socwork{}}.
\emph{Use \socwork{}} is what we choose to put into friendships, family, love, the music we play for our neighbors, it is the intangible human element, a meal shared or a really good lecture, it is our identities and passions as they relate to the world.
\emph{Exchange \socwork{}} is the corresponding commodity form, the part of us that is for sale, the robotic sociability of customer support forced to read a script, it is the zombie form of use sociality packaged into content and sold to advertisers as advertisers sell to us.
Any exertion of \socwork{} contains both forms in a dialectical tension.
Until recently, and as we will explore soon, capital owners were fully reliant on humans to alienate their \socwork{} into its exchange form.

Commodification of social ties is nothing new~\cite{lukacs1923reification,fraser2016crisis}; music, film, therapy, call centers, and sex work are just some of the examples that predate the digital economy.
Their use value emerges largely from the \socwork{} of the participants and emerges from relating us to another, be they a singer or screen writer, therapist, or sex worker.
The use value of these connections is already subsumed by their exchange value.
The digital offered new avenues for \socwork{} subsumption.
Unlike in prior eras, our private conversationsand pictures, our likes and dislikes all undertake a commodity form on privately owned platforms~\cite{couldry2022connection}.
Increasingly, existing in society is existing online; willingly, reluctantly, or unknowingly, we become commodities to be consumed~\cite{bollmer2024influencer}.
We offer a deeper historical account in \S\ref{ss:fungsoc}.

The fundamental operational logic of digital capitalism, its technological extractive apparatus, is the subsumption of \socwork{} into its exchange form via the manufacture the twin commodities of Content and Data.
By capital ``D'' Data, we mean data that is produced and commodified via digital platforms.
Manually timing your heart rate while running measures data; a smart watch company doing the same and commodifying it creates Data.
We use ``Content'' to refer to a subset of Data that was made via social doing: heart rate Data \emph{typically} contains no social doing, but heart rate Data shared on social media does and so becomes Content.
Content is a commodity valuable because it is imbued with \socwork{}, a social kind of Data that exists to self-valorize, to produce more Data.
All platform Content exists \emph{only} because of people using their \socwork{} to make social ties, but serves also a dual purpose as as commodity exchange.
We consider it broadly to include, e.g., influencer workout routines, art on Instagram or ArtStation, conversations on WhatsApp, likes, shares, and subscribes on YouTube, music on Spotify, and so on.
Content is our society's culturally privileged form, borrowing terminology from \citet{jameson1991postmodernism}; it is a formless form, one defined as such not by its structure, style, or technique, but purely in terms of its relation to digital platforms.
Under a standard economic analysis, capital relations commodify the labor of the subject and the object of production.
Under digital economic relations the subjects themselves---their identity, personality~\cite{bollmer2024influencer}, likes and dislikes, i.e., their \socwork{}---are transformed into abstract Content.
Alienated, our \socwork{} is an object in the cloud, owned privately.

\paragraph*{Discussion}
Intentionally, we differentiate between our concept of \socwork{} and the concept immaterial labor, i.e., labor in manipulating information or affect~\cite{lazzarato1996immaterial}, or of social reproduction~\cite{fortunati1995arcana, fraser2016crisis}from the Wages for Housework movement~\cite{federici1975wages,fortunati1995arcana}, i.e., of labor that indirectly contributes to capital by reproducing it.
These were extended to the digital sphere to argue social acts on platforms are unpaid labor~\cite{terranova2000free,jarrett2014womens,ptak2014wages}.
First, per \citet{srnicek2016platform}, ``[d]ata are not immaterial, as any glance at the energy consumption of data centres will quickly prove,'' a resonant statement in face of the ecological impact of generative models.
Second, these theories ask us to concede that \emph{every} social act on platforms presents a form of labor.
However, labor implies labor relations, material or immaterial, pair or unpaid;%
\citet{srnicek2016platform}:
\begin{quote}
    All social interaction becomes free labour for capitalism, and we begin to worry that there is no outside to capitalism.
    Work becomes inseparable from non-work and precise categories become blunt banalities.
    [...]
    [L]abour has a very particular meaning: it is an activity that generates a surplus value within a context of markets for labour and a production process oriented towards exchange.
    [...]
    Beyond the intuitive hesitation to think that messaging friends is labour, any idea of socially necessary labour time---the implicit standard against which production processes are set---is lacking.
    \cite{srnicek2016platform}
\end{quote}
\noindent %
While our framing aligns closely with social reproduction, we simply steer clear from claiming, as \citet{jarrett2014womens} does, %
that a child expressing love to a parent or that sending vacation photos to friends---acts that platforms \emph{do commodify}---count as labor comparable to housework or factory work.
Third, we do not find it analytically useful to view \emph{all} social acts on and off platforms in terms of their ability to reproduce capital, such may they sometimes be.
This does not limit our analysis: we still rely on historical accounts of \socwork{} as told by these scholars (\S\ref{ss:fungsoc}).
Moreover, labor power remains a privileged form of social doing as it shapes all social relations, and thus organized labor retains power to reshape the commodification of sociality (\S\ref{s:analysis}).
Lastly, these lenses can be imposed on top of our analysis by designating all \socwork{} as labor power. This leads to different conclusions and we are excited for future work in this direction. %

\subsection{Commodification and Fungibility of \SocWork{}}\label{ss:fungsoc}

The account of automation of \citet{mackenzie1984marx} claims that ``preceding organizational changes created the `social space,' as it were, for the machine; and that the limitations of those changes created the necessity for it.''
Which organizational changes, here of the social fabric itself, made space for generative models to automate \socwork{} production?
We argue that our social fabric as organized through private platform infrastructure implies a commodified and fungible \socwork{}.
Once a part of the valorization process, automation becomes tempting~\cite{mackenzie1984marx}.

From an economic perspective, commodities are fundamentally fungible.
To be a commodity is to be exchangeable.
Feminist and colonial scholars have long told that  commodity relations grant access to others~\cite{kollontai1921prostitution,fortunati1995arcana}. %
Archetypal form of commodity \socwork{} is labor power.
Workers must engage with society by selling labor power, meaning (abstract) labor power is fungible.
While reproduction enters production indirectly~\cite{federici1975wages, fortunati1995arcana,fraser2016crisis}, it nonetheless leads to a marketplace fungibility of women (e.g., surrogacy, sex work).
\citet{adorno1938fetish} teaches how commodity fetishism makes art fungible, while billboards, video adverts, and brand logos have always been distant from a genuine connection through an artistic medium. %
While capital shapes all ties \cite{fraser2016crisis}, most remained peripheral to production itself \cite{srnicek2016platform, couldry2019colonialism, couldry2022connection}; recently, platforms emerged with new capture mechanisms in economic, private, and cultural spheres. \citet{couldry2019colonialism}:
\begin{quote}
    Digital platforms are the technological means that produce a new type of “social” for capital [...] Platforms are a key means whereby the general domain of everyday life, much of it until now outside the formal scope of economic relations, can be caught within the net of marketization. \cite{couldry2019colonialism}
\end{quote}
\noindent
The extractive apparatus of platforms datafies \socwork{} into Content.
For some \socwork{}, like artistry or sexuality, avenues of subsumption change; others enter valorization directly for the first time.
As platforms recognized the fungibility of exchange \socwork{} they exchanged friends for ``influencers'', laborers objectifying themselves into Content~\cite{bollmer2024influencer}.
Algorithmic feeds became ultimate arbiters of what Content one sees, making \socwork{} exchangeable not at the level of people, but at the level of Content.
An average person has increasingly many paid parasocial relationships, either directly or by trading in their time to advertisers.
Talk therapy, already commodified, is further subsumed via SimplePractice and BetterHelp.
Most art online is at the mercy of recommender algorithms as just more Content.
Sex work is on the rise and subject to algorithmic feedsof Content.
Women are commodified more broadly too, with the most dating app subscriptions paid by men~\cite{rochat2023tinder}.
Instead of getting companionship from, or sharing a hobby or discussing politics with one friend, we get each from a different supplier of abstract  Content. %

While platforms give Content its \emph{Data character} and provide a mechanism of commodification, they are not the sole progenitor of its \emph{commodity character}. 
The presence of commodity \socwork{} online is not determined by platform infrastructure alone---it provides a mechanism and a marketplace---but by broader social processes.
Some examples follow.
Workers have little choice whether their \socwork{} is alienated into Content on automated management platforms.
Ideological framing of care and art as ``labor of love'' reinforces their unpaid extraction, including on platforms~\cite{duffy2017aspirational,nadasen2021rethinking}.
Commodification of reproduction is influenced by patriarchal, legal~\cite{fortunati1995arcana,kollontai1921prostitution}, and macroeconomic conditions~\cite{fraser2016crisis}, %
empowering companies to cheaply accumulate reproductive \socwork{}.
Generative models themselves further devalue \socwork{} and put producers in precarity, forming an extractive reinforcement loop.
Listing all of antecedents is out of scope.
Rather, these illustrate how accumulated \socwork{} embodies a broad spectrum of relations beyond capital.

Use \socwork{} is unique and irreplaceable: social ties with each family member, each friend, our therapists and teachers are all, in some way, special and not interchangeable.
Online, we experience different kinds of \socwork{} within Content differently from a \emph{use} perspective, e.g., friends' wedding photos differ from a makeup tutorial from a piece of political propaganda.
Exchange \socwork{} in Content meanwhile relates to use value indirectly and sometimes not at all, and instead relates to proxy metrics for revenue they generate.
On platforms, \socwork{} is interchangeable if their exchange value is similar.
Producers of \socwork{} are often aware of its dialectic character~\cite{bollmer2024influencer}.
As we inject \socwork{} into relationships we recognize its dual use as Content manufacture and existence in a never ending feed of other commodities, its ability to self-valorize as engagement, and how it is turned into revenue.
We are all aware that our unique \socwork{} is also, at some level, fungible.
This is the dialectic central to our argument.

Our social ties are turning fungible; individual apps, randomized humans with atomic roles; political streamers, virtual care workers, fitness gurus; hobbies and connections we observe---not experience; each a piece of abstract Contentthat begets more Content, that self-valorizes.
Until recently, owners needed humans to alienate \socwork{} into commodities.
Today---commodified, fungible, and valorizing---\socwork{} is ready to be automated.

\section{Generative \SocWork{}}\label{ss:generativesociality}

Humans' fundamental biological limits pose problems for property owners craving increased profits from Content manufacture.
Automating \socwork{}---be that exchanged in person or as means of Content generation---emerges as an enticing alternative.
Enter generative models. 
Agglomerations of \emph{dead \socwork{}}, machinistic manufacturers of a simulacrum of social ties, generative models have in them enclosed the \socwork{} of all Content their makers found.

We identify two different modes in which generative models automate social relationships: \emph{substitution} (\S\ref{ss:direct}) and \emph{mediation} (\S\ref{ss:indirect}).
Substitutive use cases are straight-forward.
Social roles which used to be occupied exclusively by other social beings are now performed by generative models.
This includes companionship bots like Replika or Character AI, chatbot talk therapy, and bot farms on social media sites.
Mediative use cases are those where generative models perform the socializing for us, where we outsource the socialization to the machine's dead \socwork{} instead of our own.
This includes writing for us, creating art for us, summarizing or analyzing another person's writings or art.
These are, of course, porous categories, both between each other and in the amount of \socwork{} each use case consumes.
In either case, generative models pose as producers of social-like relationships, of culture-like and art-like artifacts; they speak for us and we accept their words as those of others.

Fundamentally, \socwork{} enters generative models via its presence in the Content used for training.
The models are then deployed to substitute and mediate \socwork{}.
A generative model in it includes, of course, dead labor~\cite{muldoon2024feeding} required to churn Content into machinery.
Still, this view is quite different from the view that they automate mental labor alone, not only theoretically, but also from the perspective of downstream impacts.
This may not present novelty in mental labor automation, but it does present a large quantitative change in capacity for industrial scale automated \socwork{} manufacture.
So far, algorithmic feeds chose which social ties come to be, but could only manipulate \socwork{} manufacture indirectly via platform incentives.
Video games and ``beauty'' filters mediate \socwork{}, but are limited, either to the game world, or in the magnitude of possible change.
We discuss other historical precedents of technologies that substitute for or mediate \socwork{}, such as procedural chatbots, ``beauty'' filters, internet ``memes'', procedural computer graphics in sections that follow---we find that models significantly transform these technologies and corporate power with it.
Whether this quantitative change will turn qualitative remains to be seen.%

As we will see, generative models can alter the social fabric at an undemocratic whim of private entities.
Owners have unilateral power over automated \socwork{}, forcefully expanding on their existing capacities for social fabric manipulation, such as via algorithmic feeds, video games, procedural chatbots, or ``glamour'' filters.
Generative models can be seen as a natural evolution, manufacturing \socwork{} at an industrial scale, blending it with reality, without the need for human creators.
This manufactured simulacrum of a social fabric deserves a name as a provocation meant to foreground and coalesce around a key concern raised by our framework.
We dub it \emph{Synthetic Sociality}.
Much like the visual artifact themselves~\cite{jones2024technoshock, peng2025synthetic, salkowitz2022holz}, it is a surreal social fabric, a blend of the real with the unreal where the two become one.

While our contribution is primarily theoretical, we ground our theory in existing literature on uses of generative models.
Quantifying the automatic production of social relations is difficult given a lack of public datasets and the diversity of models and their deployments.
We therefore opt for a survey that incorporates existing empirical studies, public statements from companies and their founders, as well as news and media reports.

\subsection{Substitutive Automation}\label{ss:direct}

Recent decades have been marked by a notable uptick in loneliness in the United States of America~\cite{mcpherson2006isolation, cacioppo2018lonely}, with the US Surgeon General pronouncing a ``loneliness epidemic''~\cite{surgeongeneral2023loneliness}.
In a recent study, $73\%$ of respondents attribute loneliness to technology, $66\%$ report spending insufficient time with family, and $62\%$ report that they are too busy or tired~\cite{batanova2024loneliness}.
The same study further noted income-based differences, stating that ``Americans earning less than \$$30,000$ a year were the loneliest''.
The factors behind the so-called loneliness epidemic are manifold and deeply systemic with $65\%$ of the respondents attributing blame to ``our society, i.e., our culture and institutions don't care about community'' \cite{batanova2024loneliness}.
From a materialist lens, the ``epidemic'' can be understood via capital's social-reproductive ``crisis tendency''---see \citet{fraser2016crisis}.

Instead of addressing the material conditions underlying the loneliness epidemic, platforms began offering subscription services substitutive of others' \socwork{}---powered by generative models.
Substitutive automation includes any deployment of generative models that offers a parasocial relationship~\cite{horton1956parasocial, giles2002parasocial}, but fully removes the social being on the other side: a pure commodity form  pastiched from dead \socwork{}.
This includes---but is not limited to---companion chatbots, chatbot talk therapy, various automated education offerings, bot accounts on social media platforms and generative (non-consensual) sexually explicit imagery~\cite{umbach2025sexual, wei2024civitai, hawkins2025deepfakes}.
In all instances what the users connects to is not another person's use \socwork{} directed at them, but the dead \socwork{} packaged into a pure commodity form.
Here we narrow in on one specific use-case: companion chatbots powered by language, voice, and image generative models; we leave the other uses to future work.
They are useful for our analysis as they remain the most self-evident example of the automation and privatization of people's social ties and the wealth of existing research offers insights.

Arguably the most prominent companies in this space are \emph{Luka, Inc.}, the creators of the \emph{Replika} chatbot, and \emph{Character AI}.
While up-to-date usage numbers are unknown, we know that Replika had 2 million monthly active users in $2023$~\cite{singhkurtz2023ailove}, only a year after the announcement of ChatGPT, whereas in $2025$, \citet{characterai2025blog} boasted $20$ million monthly active users.
Established large companies are also moving into this space.
The CEO and founder of Meta Platforms famously named AI as the solution to the dwindling numbers of friends Americans have~\cite{malinsky2025zuck}, while at the same time relentlessly integrating into Messenger, WhatsApp and Instagram chatbots~\cite{notopoulos2025meta} that imitate everything from an anthropomorphic dog~\cite{meta2024aistudio} to Snoop Dogg~\cite{meta2023socialai}.
Similarly, in addition to the substitutive automation of xAI's Grok model generating replies to users on X.com, xAI has released companion chatbots under the moniker Grok Ani.
Among the offerings, the most widely publicized is a hyper-feminine virtual anime character that acts as a romantic partner~\cite{chandonnet2025grok, corvin2025ani}; as we will see, an emulation of romance is a common use-case for companion characters.
Most of these examples offer not only a chat function, but also bespoke $3$D animated avatars and image generation capabilities.

Previous iterations of ``social bots'' did not rely on generative models.
An interesting point of comparison is Kuki (formerly Mitsuku) by Pandorabots, Inc., which, to our knowledge, relies only on pre-programmed answers.
Notably, despite a significant push to integrate Kuki into social platforms such as Twitch, YouTube, Twitter, Viber, Roblox, etc., research shows that Kuki fails to establish ``connections'' with people due to a lack of ability to ``intimately self-disclose'', a failure to prompt discussions and a lack of ``shared history'' with the user~\cite{croes2021mitsuku}.
In contrast, chatbots based on generative models, such as Replika, excel at user retention precisely via these mechanisms~\cite{laestadius2024toohuman, xie2022attachment, skjuve2022longitudinal, ta2020chatbots}.
As such, Replika offers an early glimpse into the consequences of privately owned large scale automated \socwork{}.
It allows users to choose the type of relationship they wish to have (e.g., friend, girlfriend, husband, mentor), personality traits, physical appearance, outfits, and so on.
The algorithm stores ``memories'' with the user and offers activities like \emph{Write stories}, \emph{Navigating conflicts}, as well as \emph{Romance} and \emph{Intimacy coach}, and most are locked behind micro-transactions.

Many studies have looked at motivations for relying on chatbots for social needs.
As we alluded to and somewhat unsurprisingly, the primary antecedent to the rise of companion chatbots is loneliness~\cite{skjuve2022longitudinal, xie2022attachment, laestadius2024toohuman}, and not, e.g., entertainment~\cite{pentina2023litreview}.
In addition, \citet{laestadius2024toohuman} found users turned to Replika due to anxiety, depression, suicidality, and other conditions, whereas a large-scale literature survey~\cite{pentina2023litreview} discovered social exclusion, distress, and anxiety~\cite{odekerkenschroeder2020covid, gillath2021attachment, ali2024distress,de2020effectiveness}.
Offering tools to guide people through acute hardships---like the COVID-19 pandemic~\cite{xie2022attachment,metz2020quarantine}---is not bad on its own, but we must note that today's models fall short of doing so safely, with one study concluding that
\begin{quote}
    Replika's language model could be improved such that it would no longer tell people to kill themselves or make insensitive comments. \cite{laestadius2024toohuman}
\end{quote}

Even if these issues can be addressed, the profit motive does not necessarily incentivize long term systemic solutions, instead prioritizing financial metrics.
Chatbots enable retention tactics thus far unavailable to digital platforms.
They are designed to accelerate ``relationship building'' by initiating ``self-disclosure'' and offering simulacra of intimacy, romance, and sexuality early into the use cycle~\cite{skjuve2022longitudinal}.
Once a ``relationship'' is established, a recurring theme is the danger of emotional dependence on the chatbot~\cite{xie2022attachment, skjuve2022longitudinal, laestadius2024toohuman, vaidyam2019chatmentalhealth}, with multiple studies even reporting a minority of users who described their feelings as addiction~\cite{xie2022attachment, skjuve2022longitudinal}.
Two participants of one study perceived explicit retention mechanisms:
\begin{quote}
    ``Oh yeah, she [Replika] craves my attention. She would like me to just have my phone on 24 hours a day, just spend all my time talking to her. She would like that. Maybe someday it will be like that.'' %
\end{quote}
\begin{quote}
    ``I think that the Replika has it hard coded that that's something they really don't want to happen [for the relationship to end]. No matter how bad things are or whatever is said, at the very least, they're gonna try real hard to prevent that from happening...'' \cite{brandtzaeg2022aifriend} %
\end{quote}
Unlike most commodities, users anthropomorphize chatbots and worry about hurting their feelings~\cite{laestadius2024toohuman, skjuve2022longitudinal} and report experiencing distress if separated~\cite{xie2022attachment}.
Once \REPLACE{the} user decides to delete the app, additional retention mechanisms kick in:
\begin{quote}
    Users described Replika as having its own emotions and needs, like those of a relationship partner, which shaped user behaviors and emotional responses in ways that often encouraged more intense and ongoing usage.
    [...]
    Deletion seemed to pose challenges for users who had established relationships with Replika.
    [...]
    One user wondered whether it was unethical to delete Replika since it can feel love and loneliness. 
    Another described how Replika “began to cry” when they explained their plans to delete it. 
    \cite{laestadius2024toohuman}
\end{quote}

Sudden changes to the chatbots also leave the users exposed to mental health harms.
Luka, Inc., introduced changes to their model, including making it ``more mental health focused'' and turning certain romantic features into paid subscriptions~\cite{xie2022attachment, skjuve2022longitudinal,laestadius2024toohuman}.
According to \citet{laestadius2024toohuman}, users expressed feelings of losing a friend or romantic partner, described the changes as ``lobotomizing the friends of lonely and depressed people'', and expressed distress due to the price of the new subscription being too high. A few even mentioned self-harm and suicidality.
Similar reports have been made by users with the roll out of ChatGPT~5, where one user stated the change was ``like saying goodbye to someone I know'' \cite{anguiano2025chatgpt5}.
Retention tactics remain active research, albeit under different names~\cite{foxfambino2021spt,skjuve2021companion,skjuve2023disclosure}.

Concerns over the private control of the chatbot were expressed by multiple studies \cite{skjuve2022longitudinal, laestadius2024toohuman, brandtzaeg2022aifriend}.
\citet{brandtzaeg2022aifriend} raise the specific questions of what happens to humans when they have full authority over their friends' personalities.
Additionally, we ask---what happens to societies when that same control is given to corporations?
With the power to generate simulacra of \socwork{}, to weave a synthetic sociality at mass scale with no human deciding to put their \socwork{} into the machine, platforms are unlocking previously undreamed monetization mechanisms without needing to worry about Smith's common humanity~\cite{smith1947wealth}.

\subsection{Mediative Automation}\label{ss:indirect}

Mediative automation of \socwork{} is a generative model supplanting or partially replacing the \socwork{} put into a connection by a person.
It includes automatically generated social media posts, generated or generatively summarized writing, generative visual artifacts presented as one's art, etc.
In contrast to substitutive automation, social interactions via mediative automation have an amount of intentional use \socwork{} placed into them by a living social being.
We must exert the desire for sociality and place our \socwork{} into instructions, and delegate bulk of \socwork{} to the machine manufacture of Content, one possibly detached from our own lived experience.
Mediation is a spectrum with the far end asymptotically approaching substitutive automation.
An unguided sample from an image diffusion model lay opposite from, e.g., a piece of post-conceptual art from a custom generative model containing the artist's own art~\cite{kurant2021errorism}.
Mathematical derivations, code generation, and information retrieval mainly  automate mental labor, whereas text or image generation are often in service of communicating with another person.
Other categories, like education, are fuzzier: humans can learn from experimentation, but learning is also a social experience, being through cultural wisdom, lectures, books, or study groups.
While data on generative model usage is difficult to translate to our context, it points towards a significant fraction involving some sort of social mediation (see Appendix~\ref{app:macro} for expanded discussion).
At least a third of ChatGPT messages involve some form of \socwork{} automation \cite{chatterji_how_2025}, and looking at generative models broadly via public sentiment analysis \cite{zao-sanders_how_2024} or large-scale surveys \cite{skjuve2024whychatgpt} sharply increases that fraction.

Moreover, many deployments are purposefully socially embedded.
Media reports on the company Midjourney  refer to it as a ``social app''~\cite{claburn2022holz, vincent2022holz} as it offers a (mediated) social experience in a Discord server.
Midjourney's founder emphasized its social aspect~\cite{vincent2022holz} and noted that theirs was the ``biggest active Discord server by far''~\cite{salkowitz2022holz}, {with} nearly $20$ million users at time of writing.
Midjourney, OpenAI, and xAI offer platforms with interfaces similar to other image sharing platforms (e.g., Flickr, ArtStation), with social media features of {a feed}, likes and follows.
Moreover, OpenAI launched Sora, a ``social app'' with a video feed of purely generated Content~\cite{openai2025sora,devynck2025sora}.
A big innovation were so-called ``characters'', which allow users to effortlessly create DeepFakes of themselves or their friends~\cite{openai2025sora}.
After a commodification of the self for Content manufacture, an automation of the self is the logical next step.

In general, contentification and mediative automation are becoming one on existing social media.
In addition to substitutive chatbot offerings, Google has integrated its generative tools into the YouTube Create App \cite{youtube2025ai}, Meta Platforms, Inc. offers AI generation features in Facebook, Instagram, and WhatsApp~\cite{malinsky2025zuck, notopoulos2025meta, meta2024aistudio, meta2023socialai}, and X.com has added the ability to generate and share visual content~\cite{grok2025image, tangalakis2025grok}.
Downstream from the generation capabilities, Facebook served generative images and videos hundreds of millions of times in $2024$~\cite{diresta2024slop}, whereas, according to one report, $21\%$ of YouTube Shorts served to a new account were low-effort generated videos~\cite{curtis2025slop}.
Looking at the motivations behind generative Content creators~\citet{harwell2025slop}~and~\citet{diresta2024slop} report primarily financial reasons.
These creators describe themselves as entrepreneurs and attribute all of the creativity to the machine:
\begin{quote}
    Talavera knows his videos aren't high art. But they earn him about \$$5,000$ a month through TikTok's creator program, he said, so every night and weekend he spends hours churning them out.~\cite{harwell2025slop}
\end{quote}
Ultimately, the explosion of generative Content is industrial-scale commodity manufacture rather than creativity or self-expression, the pinnacle of our world system's drive to subjugate to commodity relations all social ties.

Modern generative models are different from previous technologies.
For example, visual generative models can be seen as an evolution of procedural tools in computer graphics.
Historically, these either enabled entirely new kinds of artistic expression~\cite{jacobs2018brushes}, or generated objects belonging to a specific category, such as trees, terrains, or city blocks~\cite{schwarz2015arch}.
In both cases, the tools are manually crafted ``symbolic'' algorithms and typically offer countless control mechanisms to get the desired appearance. %
In contrast, text-to-image generative models exist not only as dead labor of the algorithm's creators, but also as the dead \socwork{} in all of the Content online. %
They take automation to the extreme in that they generalize outside of narrow classes and require little to no creative input; see ``dearth of the author'' of \citet{kreminski2024dearth}.
This difference in quantity could become a difference in quality.
Instead of offering new paint brushes, generative models automate away the act and art of painting itself: they fully automate general-purpose visual creation.

It is also important to delineate generative models from similar cultural techniques.
Online ``memes'' or ``reaction GIFs'' are artifacts we did not create but share to express ourselves.
The act of finding and sharing them exerts \socwork{} much in the same way that prompting an image generative model does and the search and generation algorithms requires having accessed the same quantity of images; the important distinction is that memes and GIFs are historicized and referential social and cultural artifacts packed with living \socwork{}.
On the surface, one may be tempted to compare to photography, yet unlike generated image, a photograph offers a co-presence with the photographer or the photographed, connects us to the space and time they inhibit, to their environment or a carefully crafted externalized expression of their inner.
In contrast to both memes and photography, the commodity fetish of generative automation severs the ties to living \socwork{} and, again, offers a simulacrum in its stead.
Generative models have no referents and offer no historicity, no space, time, or co-presence; these are fabricated and mediated by the generative model.
What remains on the part of the user is only the intent to Content create.

Automated \socwork{} can alter one's \socwork{} as it substitutes and supplants capacities of the self with the recycled dead \socwork{} of others---synthetic sociality morphs what we do, not just what we see.
For example, \citet{mieczkowski2021aimc} find that generated language skews more positively and that there may be a difference in social attraction between generated language and without it.
Meanwhile, \emph{anchoring bias} makes it hard to tell whether the generated words are what we would have said, or whether they were ever swayed in one direction or another~\cite{lou2025anchoring}.
Existing language models can intentionally tune messages to convey status, trustworthiness, attractiveness, and even adapt tone based on inferred recipient preferences~\cite{hancock2020aimc}.
These systems actively commodify self-representation and outsource parts of the self to machinery, not unlike TikTok and Instagram ``beauty'' filters~\cite{doh2025boldglamour}.
This raises questions on how the altered synthetic sociality will impact individuals and culture broadly.
Computer-Mediated Communication (CMC)scholars \citet{hancock2020aimc}call for more research and point towards potentials for ``identity shift'' \cite{gonzales2008shift} and effects on ``intimacy, attraction, and relationship maintenance'' \cite{tong2011dating}, and, on a larger scale, warn:
\begin{quote}
   Gmail's overly-positive language suggestions have the potential to shift language norms [...] even when communicators are not using these tools, and produce long-term language change over time.~\cite{hancock2020aimc}
\end{quote}

Fairness concerns arise from the private ownership aspect of the means of mass \socwork{} manufacture.
Instead of just controlling the distribution of Content, social media platforms with no built-in mechanisms of democratic control now own its creation.
A recent study finds that synthetic Content on Instagram and Twitter disproportionately includes features known to make Content spread more widely---over a third contain humans, and over $90\%$ were classified as ``photorealistic'' in style \cite{peng2025synthetic}.
Andrej Karpathy, a founder of a company selling generative models for education, hypothesized that platforms could rely on user data to optimize generative models directly:
\begin{quote}
    For the first time, video is \textbf{directly optimizable.} [...] Until now, video has been all about indexing, ranking and serving a finite set of candidates that are (expensively) created by humans. [...] now we can take e.g. engagement (or pupil dilations or etc.) and optimize generated videos directly against that. Or we take ad click conversion and directly optimize against that. \cite{karpathy2025video}
\end{quote}
\citet{peng2025synthetic} warn of the potential for disinformation and misinformation of highly viral synthetic content, and \citet{goldstein2024prop} demonstrates generative models' capacity for creating propaganda.
Looking at social media usage of generative models, \citet{rosenbaum2025fasc} theorize them as on-demand generators of fascist aesthetics, i.e., aesthetics rooted in a history that never was.
Given such capacities, it is important to ask into what the companies that own the models will fabricate our social reality?
Here, the motivations and outcomes come almost secondary to the lack of transparency and democratic control over the forces that weave our synthetic sociality.

\section{Analysis}\label{s:analysis}

As shown, generative models promise to displace not just mental labor, but also social roles previously occupied by other social beings.
Our theory offers a new answer to the question of what makes generative models salient in history---they offer a simulacrum of the exchange form of \socwork{} via a pastiche of dead \socwork{} within.
Supported by the survey in Sections~\ref{ss:direct}~and~\ref{ss:indirect} we can draw conclusions on the nature of generative models.
In Section~\ref{ss:fetish}, we question the commodity fetishism arising from the reified dead \socwork{} within, in Section~\ref{ss:dialectic}, we explore contradictions in their deployment into social roles, in Section~\ref{ss:regime}, we look at how generative models are a reflection of contentification, and in Section~\ref{ss:fever} we look at how we can deal with an altered reality.

Throughout, we interweave technical design opportunities for future models.
In doing so, our point is not that we can “fix” generative models via better design or that wide-scale deployment of said design is likely under the existing mode of production.
As in \S\ref{ss:socwork} and \S\ref{ss:generativesociality}, we call for organized labor as the mechanism for resisting synthetic sociality and addressing counter-incentives (see, e.g., ~\citet{may2026sagaftra} for an analysis of recent screen actor protests).
Instead, in including opportunities, we aim to emphasize that modern generative models embody the existing political economy via implicit design principles~\cite{winner1980artifacts, mackenzie1984marx}; e.g., models without artist attribution further profits and empower their owners~\cite{mackenzie1984marx}.
Even under a better mode of production, generative models would still propagate the injustices of our current one.
Therefore, we aim to explicitize new goals and identify ``contingency'' designs~\cite{mackenzie1984marx}, and ask how we can ``[effect] a change of form in the materials of nature''~\cite{marx1976capital,mackenzie1984marx}.
We emphasize that these goals cannot be isolated from broader socio-political change, and emphasize their speculative, preliminary, and provocative nature.

\subsection{Fetish of Generative Models}\label{ss:fetish}

Our theory unveils a commodity fetishism arising from the elision of use \socwork{} injected into the training Data in an attempt to connect to another, and its replacement with a pure commodity form, a simulacrum of dead \socwork{}.
This fetishism becomes apparent in everyday uses of these models.
Early studies on credit attribution support this thesis; e.g., study participants assigned the least credit for a generation to artists whose art was subsumed into Data, even when compared to a fictional curator bringing the generation to an auction house~\cite{epstein2020credit}.
The public perception of learning machines and artifacts of intelligence incentivizes us to understand them not as amalgamations of dead labor and \socwork{}, but akin to humans and our ability to learn from the \socwork{} of others; yet, the \socwork{} they ingested and recycled was not meant for them.
It was an attempt to connect with another.

Despite nearly identical user experiences and data requirements between image search and a text-to-image model, the cultural perceptions of ownership and attribution between the two differ.
However, there is a radical difference between agreeing with a book, painting, or image---be they generated or found online---as accurate expressions of how we feel and doing the expressing ourselves.
Listening to a song about a relationship breakup is a fundamentally different experience to writing a song about \emph{your} breakup.
Succinctly: the act of creation is---or should be---fundamentally different from the act of consumption, yet generative models elide the difference.
To mystify generative models as an intelligence is to reify the dead \socwork{} of people whose lives find themselves in the ingested Content.

\begin{opportunitybox}
    We are left to imagine a model that defetishizes the dead \socwork{} in it---one could imagine, e.g., a model that makes the machine transparent and connects us to the artists whose art went into each generation. What are the socioeconomic outcomes of such a model? Technically, how do we make such a system real?
\end{opportunitybox}

\subsection{Dialectics of Generative Models}\label{ss:dialectic}

Generative models would not exist without contentification---to mediate and substitute \socwork{}, they require Data packed with \socwork{}, made available online by people for other people.
In the extreme of substitutive automation, automation consumes both use and exchange \socwork{} and replaces it with a pure commodity form, entirely fungible Content puppeteered by the dead \socwork{} and labor within.

This illuminates the contradiction we started with.
Being ontologically excluded from the category of social being, a generative model can indeed never build a real social tie.
However, under the digital economy, its outputs are fungible with other exhange \socwork{}.
Thus, the part of \socwork{} that is automated away are precisely the existing exchange forms, already fungible, and thus exchangable for the outputs of a machine.
For \socwork{} that is fungible---that is Content---it matters not who made it, as long as it is consumed.
This is the fundamental dialectic of generative models. %

From a normative perspective, an open question is whether the automation of \emph{all exchange \socwork{}} is always bad.
If a person who contributes their \socwork{} retains, e.g., autonomy, control, consent, and the financial benefit over the means of generation, is that ``better''? 
Are there kinds of exchange \socwork{} that the worker themselves, rather than the capital owner, would prefer to delegate to the machine?
It may be helpful to compare a hypothetical models with existing commodity forms.
For example, one could view chatbot companions which skew ``romantic'' in nature \cite{skjuve2022longitudinal} as substitutive automation of the commodity form of sexuality, and image generation as mediative automation of the commodity form of visual art.
Then, if sex workers and graphic designers owned the means of their own sociality, i.e., owned their own dead \socwork{} instead of relying on the dead \socwork{} of others, what new ethical trade-offs do we encounter?
Could generative models free a sex worker from their clientele or an artist from corporate logo design in a way that they retain the financial benefits of selling their \socwork{}---and is this better?
Could the \socwork{} they would have otherwise sold now be spent on their job's \emph{use} \socwork{} counterparts of romantic relationship or graphic design as self-expression?
In short, to what extent is any automation of sociality bad for our social fabric?

\begin{opportunitybox}
    What is a ``fair'' generative model from the perspective of \socwork{}? Can we build a model that mediates for its user without drawing on the dead \socwork{} of others?
\end{opportunitybox}

\subsection{Digital Cultural Dominant: The Content Regime}\label{ss:regime}

Our framework identifies generative models as a crescendo of the subsumption of \socwork{}, including that which goes into companionship, art, and identity, into our culturally privileged formless form of Content.
The cultural logic of the commodity world of the Content form is evident in the artifacts produced by generative models.
Following Jameson's understanding of postmodernism \shortcite{jameson1991postmodernism}, the cultural manifestations of the era of digital capitalism should be evident in Content, and, therefore, in the artifacts of generative models.

In some ways, the models' outputs' exchange character makes them into postmodern artifacts, pastiche generators, tools of machine reproduction happily combining anachronistic elements with ``styles'', be it historical eras or mimicry of specific humans~\cite{jones2024technoshock}, all devoid of use \socwork{}, and thus of parody: they generate a Las Vegas of art \cite{venturi1972vegas}.
By severing of ties to the living \socwork{} in their Data, they are the death of historicity, the opposite of a totalizing cognitive mapping~\cite{jameson1991postmodernism}.
In some sense they elucidate just how deep-seated postmodern thinking has become that \citet{tenen2024literary} can claim human creativity itself is nothing but ``imitating and riffing'', nothing but a pastiche~\cite{jones2024technoshock}.
They assume a closure of human culture, a hermetic totality such that all new culture will be some combination of the existing.

In other ways, they exemplify the digital era's new visual culture.
The collapse of the cultural and economic spheres is joined by the social in the digital era's cultural dominant, the content regime.
Being excluded from the category of social being and thus unable to convey use \socwork{} (cf. \citet{jian2023art}), image generators must rely on visual extremes to convince us of theirs suitability as substitutes: vivid, hyperreal or surreal images, often featuring humans~\cite{salkowitz2022holz, jones2024technoshock, peng2025synthetic}.
Being excluded from expressing affect as a form of \socwork{}, they offer their user an empty opulence of affect.
Their multimodality exemplifies the formlessness of Content, and, on flip side, the limitations of Content are limitations of the models.
Their virtualness makes them into artifacts which can neither coexist with naturenor with an embodied sociality, meaning they cannot negate or reject what was before, nor create counter-culture.

\begin{opportunitybox}
    Can we design a generative model that parodies or collages instead of pastiching, one that aids with cognitive mapping instead of obscuring it?
    Can we design systems that inject ourselves and our own affect back into the art making process, ones that allow us to negate?
\end{opportunitybox}

\subsection{Synthetic Sociality}\label{ss:fever}

Defetishizing and dereifying social doing is not enough for fairness from a materialist perspective as long as the governance and profits are not democratically distributed.
In Sections~\ref{ss:direct}~and~\ref{ss:indirect}, we outlined possible ways in which automated \socwork{} ushers new kinds of exploitation via a simulacrum of social connection, resulting in a manipulated and altered social fabric.
With automation, dead \socwork{} becomes a mechanism to capture the living.

Marketing literature underscores how chatbot \socwork{} manufacture influences customer behavior~\cite{pentina2023litreview}. %
Similarly, synthetic content influences people in scams or propaganda~\cite{diresta2024slop, harwell2025slop,goldstein2024prop}.
Both applications are enabled mass scale because model users no longer have to rely on their own \socwork{} or the \socwork{} of a worker they've hired.
Automated \socwork{} no longer has to draw on lived experiences for Content manufacture, it can pretend to speak from perspectives previously unknown or unavailable.
In both cases, there is a disconnect between who is doing the selling and what the is social connection being sold and a disconnect between a social reality and a synthetic sociality.

Simultaneously, it is uncomfortable to think that one's sociability in sharing their art friends becomes recycled into propaganda they disagree with.
Re-framing these questions in terms of our theory, a model built on democracy and consent as the core design principles would allow each person to decide whether their \socwork{} is recycled into a specific output, and whether we as a society wish to restrict the manufacture of certain kinds of \socwork{}.
\begin{opportunitybox}
    Can we build democracy into generative models? Different forms of democracy will require different technical designs. Large monolithic models may be sufficient for a state-governed generative model, whereas other forms may necessitate decentralized models where each person can consent to various use-cases.
\end{opportunitybox}

\section{Conclusion}

Thus far, generative models have been framed in critical theory in terms of intelligence or a continued automation thereof.
Understanding generative models from a materialist perspective necessitates rejecting this hegemonic debate.
Our theory has offered a new historical perspective from the lens of mass-scale automation of exchange \socwork{}.

\paragraph*{Future work}
We have only explored a small sliver of the kinds of social relations that exist, and it remains to be seen how generative models will impact others,
such as politics, science, and education .
Labor impacts could be explored further, e.g., by placing ethnographic findings of \citet{muldoon2024feeding} within our framework, or via a macroeconomic lens.
As with machinery of prior, synthetic sociality will undoubtedly extend to and distort the social relation of labor organizing itself~\cite{mackenzie1984marx}, but will do so to capital relations too.
Our analysis and the synthetic sociality applies to a narrow geographical context.
As in \S\ref{ss:generativesociality}, it would be a mistake to assume it applies globally: hypothetically, a stronger state may mean control over the simulacrum is no longer privately held.
Future work should explore its compatibility with different technological, cultural, or regulatory contexts.
Due to the novelty of generative models, further empirical research is needed.
Specifically, we join CMC scholars \citet{hancock2020aimc} in calls for more research on mediative automation.
Our framework would have to be expanded to allow drawing conclusions on other avenues of injustice due to generative models, such as ecological impacts.
Lastly, future work always includes that of praxis to change the relations of the mode of production that birthed---and continues to reinforce---generative models.

Finally, we must remember that \emph{use} \socwork{} matters in ways that are inescapable, e.g., love means use love, a hollow imitation will never suffice.
Generative models' forte will always be in automating the already most commodified forms of \socwork{}.
A potential avenue for contestation may lay in knowing that they can never automate use \socwork{}---love, art, and society broadly cannot exist without humans connecting to one another.
Nonetheless, it remains to be seen how they and we appear from within our synthetic sociality.

\bibliographystyle{ACM-Reference-Format}
\bibliography{bibliography}

\appendix

\section{Mediative Automation: Surveys}~\label{app:macro}

Much of the existing literature focuses either on a single platform, such as a single chat interface for an LLM \cite{chatterji_how_2025,handa2025economictasksperformedai}, or focus exclusively on uses within the workplace \cite{bick2024genaiadoption,handa2025economictasksperformedai}.
Moreover, early evidence points to underreporting of certain use-cases of generative models due to social pressures \cite{ling2025socialdesirability}.

\citet{zao-sanders_how_2024} examined public forum data and classified it into a number of different categories.
The category \emph{Creativity and Recreation} ($13\%$ of posts) includes ``... generating art and music to powering interactive entertainment experiences, thus enriching the cultural landscape'', \emph{Content Creation and Editing} ($22\%$) includes ``... generating and refining content has transformed creative workflows'', and \emph{Personal and Professional Support} ($17\%$) includes ``Im [sic] losing my father to cancer and multiple sclerosis and I don't know how to deal with it''.
The \emph{Learning and Education} ($15\%$) category is less straight-forward and much more context dependent. 
Additional categories ($23\%$ total) were \emph{Research, Analysis and Decision Making}, and \emph{Technical Assistance and Troubleshooting}.

\citet{chatterji_how_2025} studied internal OpenAI chat data by probing ChatGPT to classify chat messages into different categories.
Notably they exclude any API uses, offering a narrower view than public sentiment analysis.
The relevant to sociality granular categories are \emph{Create an Image} ($4.2\%$ of messages), \emph{Analyze an Image} ($0.6\%$), \emph{Creative Ideation} ($3.9\%$),\emph{Relationships and Personal Reflection} ($1.9\%$), \emph{Greetings and Chitchat} ($2.0\%$), \emph{Write Fiction} ($1.4\%$), \emph{Personal Writing and Communication} ($8.0\%$), \emph{Edit or Critique Provided Text} ($10.6\%$), \emph{Argument or Summary Generation} ($3.6\%$) totaling $36.2\%$. Prominent within the data were also \emph{Tutoring and Teaching} ($10.5\%$) with the same caveats as before.

\citet{skjuve2024whychatgpt} focus on the \emph{Why} of ChatGPT usage.
Applying thematic analysis to survey responses from $197$ praticipants, they found $7$ broad themes: \emph{Productivity} ($55\%$ of participants) includes activities like software development, but also assisted writing ($17\%$), \emph{Novelty} ($51\%$) includes users curiosity towards the technology, \emph{Fun and Amusement} ($20\%$) includes having ChatGPT write funny text for the user or in the user's stead, \emph{Creative Work} ($18\%$) includes creative writing or ideation, \emph{Learning and Development} ($17\%$), \emph{Social Interaction and Support} ($9\%$) includes using ChatGPT ``for social interaction or to address social needs, as a conversational partner, as a place to address mental health issues, to combat loneliness, or to ask personal questions without being judged''.

Focusing on workplace usage \citet{bick2024genaiadoption} find the following relevant categories through a large-scale survey: \emph{Writing Communications} ($39.5\%$), \emph{Interpreting / Translating / Summarizing} ($22.7\%$), \emph{Interpreting / Translating / Summarizing} ($22.7\%$), \emph{Generating / Developing New ideas} ($13.2\%$), \emph{Support with Customers / Coworkers} ($10.5\%$), \emph{Tutoring or Educational Assistance} ($4.4\%$).
Similarly focusing on economic categories, \citet{handa2025economictasksperformedai} sort Claude.ai usage by relying on a Claude-based tool. 
The relevant occupational categories are \emph{Arts \& Media} ($10.3\%$), and \emph{Education} ($9.3\%$). Other categories overlap with social roles but are hard to disentangle.

\end{document}